# Measurement of polarization quantum states under chromatic aberration conditions

Yu.I. Bogdanov[*], B.I. Bantysh, N.A. Bogdanova, M. I. Shakirov, V.F. Lukichev

*Valiev Institute of Physics and Technology of Russian Academy of Sciences, Moscow, Russia*

## ABSTRACT

The wave plate is a basic device for transforming and measuring the polarization states of light. It is known that the transformation of light by means of two wave plates makes it possible to measure the state of polarization in an arbitrary basis. The finite spectral width of the light, however, leads to a chromatic aberration of the polarization quantum transformation caused by the parasitic dispersion of the birefringence of the plate material. This causes systematic errors in the tomography of quantum polarization states and significantly reduces its accuracy. This study is a development of our work[1], in which an adequate model for quantum measurements of polarization qubits under chromatic aberration was first formulated. This work includes a generalization of the results obtained earlier for the cases of two-qubit states. Along with examples of random states those uniformly distributed over the Haar measure are considered. Using a matrix of complete information, it is quantitatively traced how the presence of chromatic aberrations under conditions of a finite spectral width of light leads to the loss of information in quantum measurements. It is shown that the use of the developed model of fuzzy measurements instead of the model of standard projection measurements makes it possible to suppress systematic errors of quantum tomography even when using high-order wave plates. It turns out that the fuzzy measurement model can give a significant increase in the reconstruction accuracy compared to the standard measurement model.

## 1. INTRODUCTION

One of the main tasks of quantum technologies is the development of devices capable of processing, storing and efficiently transmitting quantum information. Optical quantum technologies based on polarization qubits are a promising platform for their implementation. Light polarization finds its application in problems of quantum communication, quantum memory, quantum computing and quantum simulators [4-7]. At all stages of development, control of quantum transformations is a very important subject. While there are various methods of reconstruction of the quantum state, quantum tomography seems to be the most reliable one [10-16].

The mathematical apparatus of polarization quantum optics is based on the quantum-mechanical description of polarization quantum states and transformations over them. Polarization state is the superposition of states of vertical and horizontal polarization of a photon: $|\psi\rangle = c_V |V\rangle + c_H |H\rangle$, where $c_V, c_H$ are probability amplitudes.

Consider generation of a pair of entangled photons through the process of spontaneous parametric down-conversion. Pump radiation decays in crystal into two modes, the signal "s" and the idle "i", respectively [3]. Figure 1 schematically shows physical implementation of the used model for measuring polarization states.

[*] bogdanov_yurii@inbox.ru

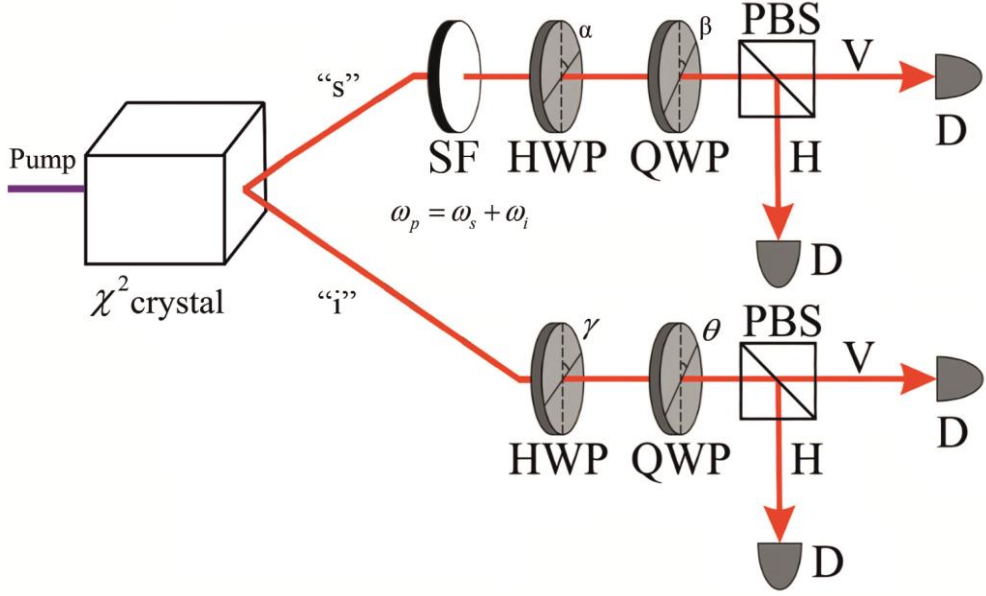

Figure 1. Polarization state measurement in an arbitrary basis: HWP- half-wave plate, QWP- quarter-wave plate, PBS- polarization beam splitter, D – photo detector, SF – spectral filter, $\chi^2$ crystal – nonlinear crystal with second-order susceptibility.

To measure polarization states of a single qubit in the $\{|V\rangle, |H\rangle\}$ basis, it is sufficient to use a polarization beam splitter and a detector in each of measurement channels. However, it is not enough to measure the quantum state in a single basis. A system of half- and quarter-wave plates is used to perform the measurements in a set of different bases [1]. The measurement model will be described in details in section 2.

The wave plate acts on the vectors of states of vertical and horizontal polarization, ensuring the rotation of the state on the Bloch sphere. It can be shown that unitary operator of the wave phase plate action is given by the following matrix:

$$U_{WP}(\delta,\alpha) = \begin{pmatrix} \cos\delta - i\sin\delta\cos 2\alpha & -i\sin\delta\sin 2\alpha \\ -i\sin\delta\sin 2\alpha & \cos\delta + i\sin\delta\cos 2\alpha \end{pmatrix} \quad (1)$$

Here $\alpha$ is the angle between the vertical and the fast optical axis, $\delta = \dfrac{\pi h |n_e - n_0|}{\lambda}$ is the optical thickness of the plate, $\lambda$ is the radiation wavelength, $h$ is the wave plate geometric thickness, $n_o$ and $n_e$ are refractive indices of ordinary and extraordinary rays in the wave plate crystal respectively.

The thicknesses for half-wave and quarter-wave plates are selected in such a way that the conditions are met at the central wavelength: $\delta_{HWP} = \pi/2 + \pi k$, $\delta_{QWP} = \pi/4 + \pi k$ (here $k$ is the wave plate order).

As the wave plate optical thickness depends on wavelength, the finite spectral width of the radiation results in chromatic aberrations. Various spectral components of the state exhibits different transformations. As a result, the output quantum state is blurred relative to the expected one [8,9]. This leads to loss of information and systematic measurement errors, which affects the accuracy of state reconstruction. In order to take into account systematic errors, we construct a fuzzy measurement model (section 3). In Section 4 we numerically compare the standard measurement model and the developed one.

## 2. MEASUREMENT MODELS FOR STATES OF ONE AND TWO POLARIZATION QUBITS

First, consider a state vector $|\psi\rangle$ of one qubit in a Hilbert space of dimension s=2. By measuring state in the $\{|V\rangle,|H\rangle\}$ basis, we obtain a binomial distribution with probabilities $p_V = |\langle V|\psi\rangle|^2$ and $p_H = |\langle H|\psi\rangle|^2$. By complementarity principle, we also need to perform additional measurements in other various bases. To do this, we subject the state $|\psi\rangle$ to some unitary transformation $U$ resulting, in the new binomial distribution with probabilities $\tilde{p}_V = |\langle V|U|\psi\rangle|^2 = |\langle \tilde{V}|\psi\rangle|^2$, $\tilde{p}_H = |\langle H|U|\psi\rangle|^2 = |\langle \tilde{H}|\psi\rangle|^2$. They correspond to measurement basis $\{|\tilde{V}\rangle = U^\dagger|V\rangle, |\tilde{H}\rangle = U^\dagger|H\rangle\}$. A set of such bases forms complementary measurements that define the quantum measurement protocol. If the protocol uniquely defines any pure or mixed quantum state, the protocol is informationally complete [13].

In our case, the unitary transformation is given by the action of wave plates. The measurement basis is changed by rotating fast axis of wave plates relative to vertical:

$$U = QWP(\delta_2,\beta) \cdot HWP(\delta_1,\alpha) = U(\delta_1,\alpha,\delta_2,\beta) \qquad (2)$$

More generally, a state can be described by a density matrix $\rho$. In this case, the probability distribution is given by

$$\tilde{p}_j = \text{Tr}(U\rho U^\dagger P_j) = \text{Tr}(\rho P_j(\alpha,\beta)), \quad j=V,H \qquad (3)$$

with projectors $P_j = |j\rangle\langle j|$, $P_j(\alpha,\beta) = U^\dagger P_j U = |\tilde{j}\rangle\langle \tilde{j}|$.

For the case of 2 qubits, a pure quantum state is $|\psi\rangle = c_{VV}|VV\rangle + c_{VH}|VH\rangle + c_{HV}|HV\rangle + c_{HH}|HH\rangle$ and the measurements are carried out in a Hilbert space of a larger dimension, s = 4.

A set of wave plates should be located in each spatial mode (Figure 1). Thus, transformation performed on the state is given by tensor product:

$$U^{(s,i)} = U\left(\delta_1^{(s)},\alpha,\delta_2^{(s)},\beta\right) \otimes U\left(\delta_1^{(i)},\gamma,\delta_2^{(i)},\theta\right) \qquad (4)$$

Similarly, we have an increasing number of protocol measurements. For example, to estimate a single qubit state using mutually unbiased bases[19], three measurements are needed. For the case of two qubits, we need to perform three measurements in the second channel for each configuration of the first channel, i.e. 3·3 = 9 measurements.

## 3. DESCRIPTION OF FUZZY MEASUREMENT OPERATORS UNDER CHROMATIC ABERRATIONS

To take into account chromatic aberrations we consider fuzzy measurements operators instead of ideal projectors $P_j(\alpha,\beta)$.

Let $P(\lambda_k)$ be the probability that photon will be inside the small spectral interval $d\lambda$ near wavelength $\lambda_k$.

By virtue of law of conservation of energy, the frequency of pump photon is equal to the sum of frequencies of signal and idle photons: $\omega_p = \omega_s + \omega_i$. We believe that pump is set by continuous laser radiation. When the pulse duration is long, the pumping frequency $\omega_p$ can be considered strictly fixed[3]. Thus, wavelength in the *"i"* mode is uniquely set by synchronism condition according to the formula:

$$\lambda_i = \frac{\lambda_s \lambda_p}{\lambda_s - \lambda_p} \qquad (5)$$

For non-monochromatic radiation, a two-mode spectral distribution $P^{(s,i)}(\lambda_s, \lambda_i) = P^{(s)}(\lambda_k)\delta(\lambda - \lambda_i)$, where $P^{(s)}(\lambda_k)$ is spectral distribution in the *"s"* mode, which is set by spectral filter (Figure 1), and $\delta(\lambda - \lambda_i)$ sets an unambiguous definition of wavelength of radiation in the *"i"* mode.

Since each wavelength corresponds to its own optical path length $\delta_k$, a chromatic aberration of the transformation occurs: $\rho \to \sum_k U_k(\alpha,\beta)\rho U_k^\dagger(\alpha,\beta) P(\lambda_k)$, where $U_k$ is the transformation (2) performed by 2 wave plates at the wavelength $\lambda_k$. The action of wave plates $U_k^{(s,i)}$ describes by (4) at the wavelength $\lambda_k$.

Using (3) and rule of the cyclic permutation of operators under trace sign, we obtain the following fuzzy measurements operators instead of ideal projectors $P_j(\alpha,\beta)$:

$$\Lambda_j(\alpha,\beta) = \sum_k U_k^{(s,i)\dagger}(\alpha,\beta) P_j(\alpha,\beta) U_k^{(s,i)}(\alpha,\beta) P(\lambda_k), \quad j = VV, VH, HV, VV.$$

## 4. SIMULATION RESULTS

To verify consistency of developed models, we consider the results of numerical simulation of quantum tomography of two-qubit states. The reconstruction of pure states is performed by means of standard and fuzzy measurement operators. Calculations are carried out for cases with finite spectral width and for the ideal monochromatic case. We will consider random pure states uniformly distributed according to the Haar measure[18].

Let the central wavelength of the "signal" channel be $\lambda_s = 0.65$ μm for a two-qubit state (e.g. see [17]). We choose the central wavelength of pump radiation $\lambda_p = 0.325$ μm. The spectral filter in the *"s"* mode cuts off the radiation spectral width to Δλ. We assume that spectrum of the "signal" channel has a uniform distribution.

We take the wave plate order to be 5, in all measurement channels. The corresponding wave plate thicknesses are $h_{HWP} = 396$μm, $h_{QWP} = 378$μm (we quartz wave plates [17]).

We consider measurements protocols based on symmetry of regular shapes, namely the cube and the octahedron ones[1].

The accuracy of pure states reconstruction is measured using fidelity $F = |\langle\varphi|\psi\rangle|^2$. It determines the probability of registering a state $|\varphi\rangle$, provided that a state is prepared in prepared $|\psi\rangle$. We also consider the loss function $L = n_{tot}\langle 1-F\rangle$, where $n_{tot}$ total sample size, and $\langle 1-F\rangle$ average infidelity. The loss function values do not depend on the sample size.

Finally, we consider the protocol efficiency $Eff = \dfrac{\langle 1-F\rangle_{min}}{\langle 1-F\rangle}$, where $\langle 1-F\rangle_{min}$ are the minimum possible average infidelity[15].

Figure 2 illustrates how finite width of photon spectrum affects the theoretical infidelity distribution[14]. One can see that with an increase of spectral width, the modes of distribution density curves become lower, and tails of distributions become longer.

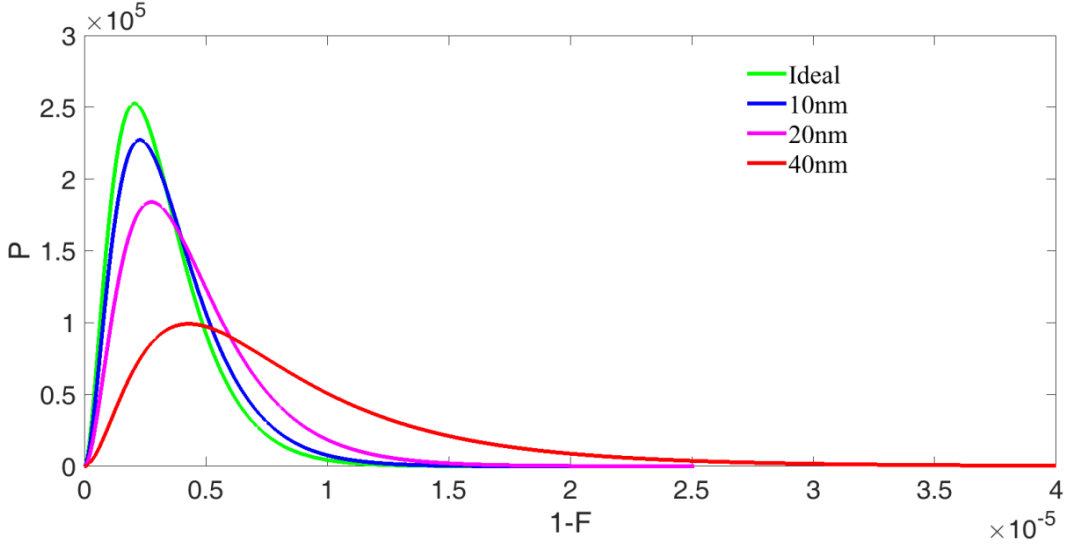

Figure 2. Averaged universal infidelity distribution for random two-qubit polarization states, for different spectrum widths Δλ = [0, 10 nm, 20 nm, 40 nm]. Protocol with octahedron symmetry in each channel; number of numerical experiments $n_{exp} = 200$; $n_{tot} = 10^6$.

Figures 3 and 4 compare protocols of a cube and an octahedron under ideal conditions when spectrum width of signal photon is vanishingly small. We see that due to larger number of measurements, octahedron protocol has slightly lower losses and slightly higher efficiency compared to the cube protocol.

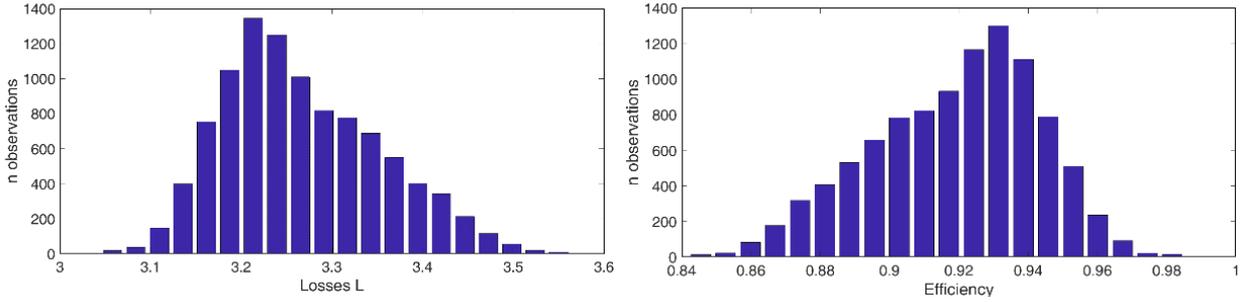

Figure 3. Distribution of Losses and Efficiency functions, an ideal case. s=4, a protocol with cube symmetry, Δλ=0, number of numerical experiments $n_{exp} = 10^4$. $n_{tot} = 10^6$; left - Losses, right - Efficiency.

$L_{mean}^{Cube,\Delta\lambda_s=0} = 3.26674 \pm 0.00087$, $Eff_{mean}^{Cube,\Delta\lambda_s=0} = 0.91899 \pm 0.00024$

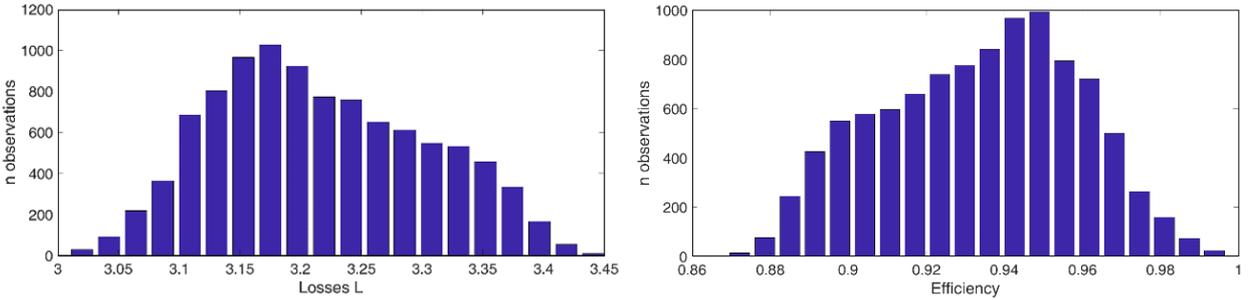

Figure 4. Distribution of Losses and Efficiency functions, an ideal case. s=4, a protocol with octahedron symmetry, Δλ=0, number of numerical experiments $n_{exp} = 10^4$. $n_{tot} = 10^6$; left - Losses, right - Efficiency.

$L_{mean}^{Octahedron,\Delta\lambda_s=0} = 3.21615 \pm 0.00087$, $Eff_{mean}^{Octahedron,\Delta\lambda_s=0} = 0.93347 \pm 0.00025$

Figures 5 and 6 compare protocols of a cube and an octahedron under conditions when width of signal photon spectrum is 20 nm. One can see that when compared with the ideal case, accuracy losses increased, and efficiency decreased by about 1.4 times. This means that to compensate for efficiency losses from decoherence caused by chromatic aberration, it is necessary to apply a model of fuzzy measurements and increase the sample size by about 1.4 times.

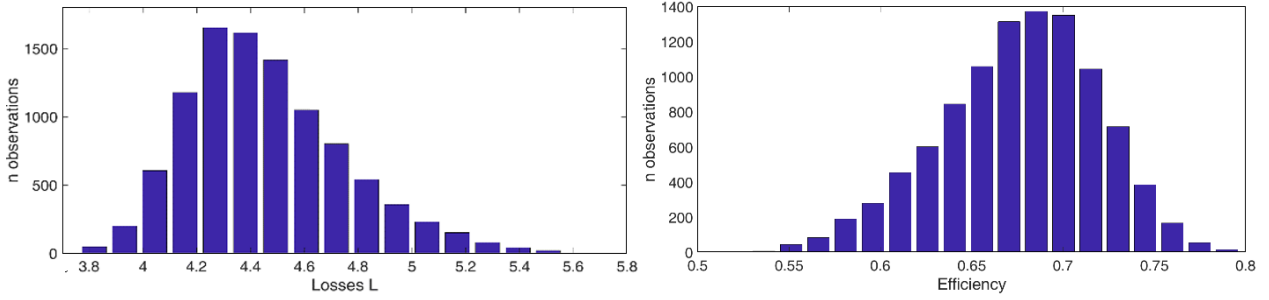

Figure 5. Distribution of Losses and Efficiency functions, final width of spectrum is 20 nm. s=4, a protocol with cube symmetry, $\Delta\lambda$=20 nm, number of numerical experiments $n_{exp} = 10^4$. $n_{tot} = 10^6$; left - Losses, right - Efficiency.

$L_{mean}^{Cube,\Delta\lambda_s=20nm} = 4.4580 \pm 0.0029$, $Eff_{mean}^{Cube,\Delta\lambda_s=20nm} = 0.67578 \pm 0.00043$

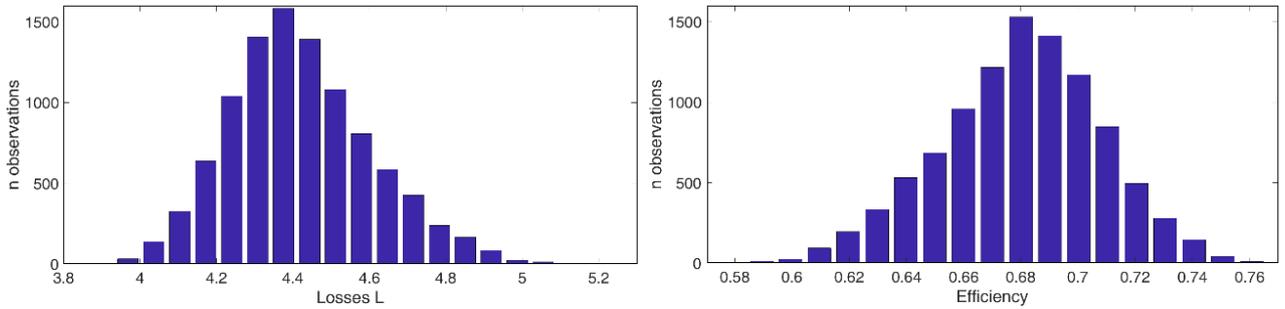

Figure 6. Distribution of Losses and Efficiency functions, final width of spectrum is 20 nm. s=4, a protocol with octahedron symmetry, $\Delta\lambda$=20 nm, number of numerical experiments $n_{exp} = 10^4$. $n_{tot} = 10^6$; left - Losses, right - Efficiency.

$L_{mean}^{Octahedron,\Delta\lambda_s=20nm} = 4.4201 \pm 0.0019$, $Eff_{mean}^{Octahedron,\Delta\lambda_s=20nm} = 0.67991 \pm 0.00028$

The use of fuzzy measurement model is fundamentally important, which is illustrated in Figure 7. Here, the reconstruction of random states using standard and fuzzy measurement models was carried out. The statistical data was generated using the fuzzy measurements operators simulating real experimental conditions.

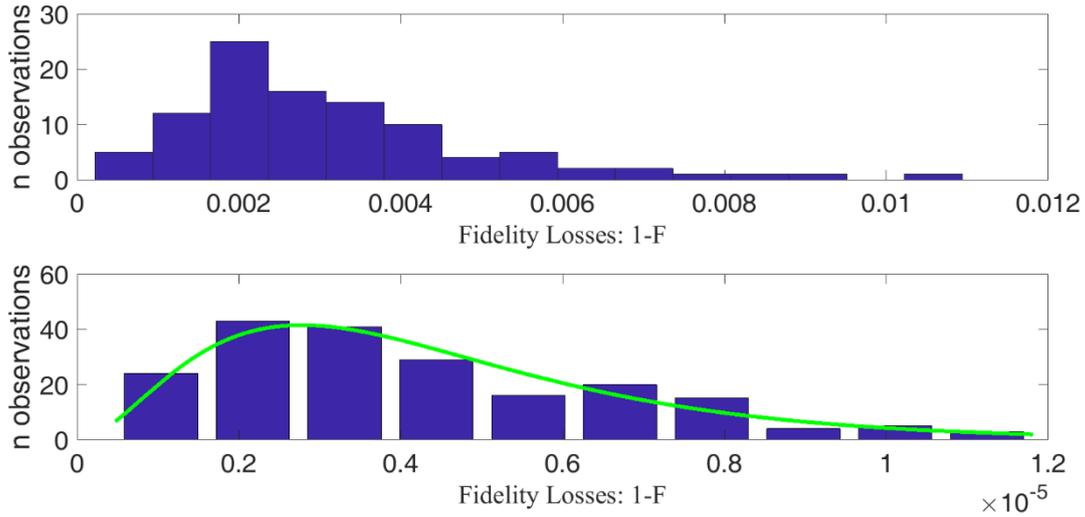

Figure 7. Comparison of accuracy losses when using standard model (top) and fuzzy measurement model (bottom). $n_{tot} = 10^6$; number of numerical experiments $n_{exp}$ =200. s=4, Protocol with octahedron symmetry, $\Delta\lambda$=20 nm.

A comparison of results of two different models shows that in this case, use of fuzzy measurement model reduced the average loss of accuracy by 738 times compared to the standard model.

The application of the chi-square criterion shows inadequacy of standard model for reconstruction of quantum states and adequacy[20] of the model developed in this study based on fuzzy measurements.

## 5. CONCLUSIONS

The spectral degree of freedom is a source of chromatic aberrations inside wave plates. Due to loss of information, tomography of polarization quantum states get worse, and the standard measurement method becomes inadequate. The use of an adequate model of fuzzy quantum measurements, taking into account chromatic aberration of light inside wave plates, makes it possible to radically increase the accuracy of reconstruction of polarization quantum states in comparison with the standard measurement model.

The developed tomography model of polarization quantum states, taking into account chromatic aberrations, is essential in tasks of high-precision control of optical quantum information technologies.

## ACKNOWLEDGMENTS

This work was supported by the Ministry of Science and Higher Education of the Russian Federation (program no. FFNN-2022-0016 for the Valiev Institute of Physics and Technology, Russian Academy of Sciences), and by the Foundation for the Advancement of Theoretical Physics and Mathematics BASIS (project no. 20-1-1-34-1)